\documentclass{kapproc} 
\usepackage{epsfig}
\begin{document}

\articletitle[Low-latitude gamma-ray sources] {Low-latitude
gamma-ray sources: correlations and variability}

\author{Gustavo E. Romero}
\affil{Instituto Argentino de Radioastronom\'{\i}a (IAR)\\ Casilla
de Correos No. 5\\ 1894 Villa Elisa\\ Buenos Aires \\ Argentina}
\email{romero@irma.iar.unlp.edu.ar}

\begin{keywords}
Gamma-ray sources, early-type stars, supernova remnants, compact
objects
\end{keywords}

\begin{abstract}
A review of the main characteristics of low-latitude sources in
the third EGRET catalog  is presented. There are 75 unidentified
gamma-ray sources detected by EGRET at $|b|<10^{\circ}$. About a
half of these sources are spatially correlated with potential
galactic gamma-ray emitters such as supernova remnants, OB
associations and early-type stars with very strong stellar winds.
The other half is formed by sources without positional correlation
with known galactic objects capable to generate a gamma ray flux
significant enough as to be detected by EGRET. A variability
analysis shows that this second group of sources contains several
objects with high levels of gamma-ray variability. These variable
sources resemble very much the AGNs detected by EGRET, but without
their typical strong radio emission. To establish the nature of
these sources is one of the most urgent problems of high-energy
astrophysics.

\end{abstract}

\section{Introduction}

The existence of a galactic population of gamma-ray sources is a
well-known fact since the early days of the COS-B experiment
(Bignami \& Hermsen 1983). The ESA COS-B satellite was launched on
August 9, 1975, and was operational until April 25, 1982. The
second COS-B catalog contains 25 sources, most of which are
located very close to the galactic plane (Swanenburg et al. 1981).
Although two low-latitude sources were soon identified with
pulsars (Crab and Vela), the nature of the remaining, presumably
galactic sources, stood uncertain. Montmerle (1979) presented the
first correlation study for these high-energy sources. He found
that about 50 \% of the unidentified COS-B detections lie in
regions containing young objects like massive stars and supernova
remnants. He estimated that the chance probability of these
associations was as low as $\sim 10^{-4}$ and suggested that the
gamma-rays could be the result of $\pi^0$-decays originated in
hadronic interactions between locally accelerated cosmic rays and
ambient gas. The cosmic rays would be produced by a two-step
process: low-energy protons or nuclei are firstly accelerated by
OB stars and injected at supernova shock fronts where they are
subsequently re-accelerated up to high-energies by Fermi
mechanism. This scenario predicted by first time the positional
correlation between low-latitude gamma-ray sources and star
forming regions, a correlation that would be tested by several
authors, using increasingly improved data, in the years to come.

A major breakthrough in the study of galactic gamma-ray sources
was achieved with the advent of NASA's Compton Gamma-Ray
Observatory (CGRO) in 1991. During its lifetime, the Energetic
Gamma-Ray Experiment Telescope (EGRET) detected 271 point sources,
170 of which have not been clearly identified yet (Hartman et al.
1999). About a half of these unidentified sources are located at
low galactic latitudes and many studies looking for correlations
with known galactic populations have been performed in recent
years. For instance, Sturner \& Dermer (1995) and Sturner et al.
(1996) have investigated the correlation between gamma-ray sources
in the first two EGRET catalogs (Thompson et al. 1995, 1996) and
supernova remnants, finding statistical support for the idea that
some remnants could be gamma-ray emitters. This idea is also
supported by several particular cases studied in detail by
Esposito et al. (1996) and Combi et al. (1998, 1999, 2001), where
the gamma-ray emission seems to come from molecular clouds
overtaken by the expanding shell of nearby remnants.

The correlation between EGRET sources and star forming regions was
confirmed by Kaaret \& Cottam (1996) and Yadigaroglu \& Romani
(1997), using data from the second EGRET catalog. Contrary to the
original hypothesis of Montmerle, these authors suggested that
most of the low-latitude sources could be pulsars. This suggestion
is supported by the discovery of several new gamma-ray pulsars
since the COS-B original identifications (there are at least seven
pulsars detected so far, see Thompson 1996) as well as by
population studies (e.g. Yadigaroglu \& Romani 1995, Zhang et al.
2000, McLaughlin \& Cordes 2000). The high level of variability
and peculiar spectral features presented by some low-latitude
unidentified sources, however, is quite at odds with the
hypothesis of a unique population of galactic gamma-ray emitters
and seem to open the possibility that our Galaxy could contain yet
unknown types of high-energy sources (McLaughlin et al. 1996,
Merck et al. 1996, Tavani et al. 1998, Tompkins 1999, Romero et
al. 2000, Punsly et al. 2000, Torres et al. 2000).

In this paper we shall review the main correlational properties of
the sample of low-latitude unidentified gamma-ray sources in the
third and final EGRET catalog. We shall discuss which objects in
our Galaxy are expected to generate strong enough gamma-ray
emission as to have been detected by EGRET and what are the
prospects for future space missions like INTEGRAL and GLAST
regarding the identification of galactic gamma-ray sources.

\section{Low-latitude sources and the spiral structure of the Galaxy}

There are 81 unidentified gamma-ray sources in the third EGRET
catalog located at less than $10^{\circ}$ from the galactic plane.
Six of these sources are thought to be artifacts associated with
the proximity of the very bright Vela pulsar. These sources do not
show up in a map which excludes the Vela pulsation intervals and
will not be taken into account in the present discussion. We have,
consequently, a sample of 75 unidentified low-latitude sources,
whose distribution with galactic latitude is shown in Figure 1. A
strong concentration around zero degrees can be clearly seen in
this histogram, indicating that most of the sources belong to our
Galaxy.

\begin{figure}[ht]
\centerline{\epsfig{file=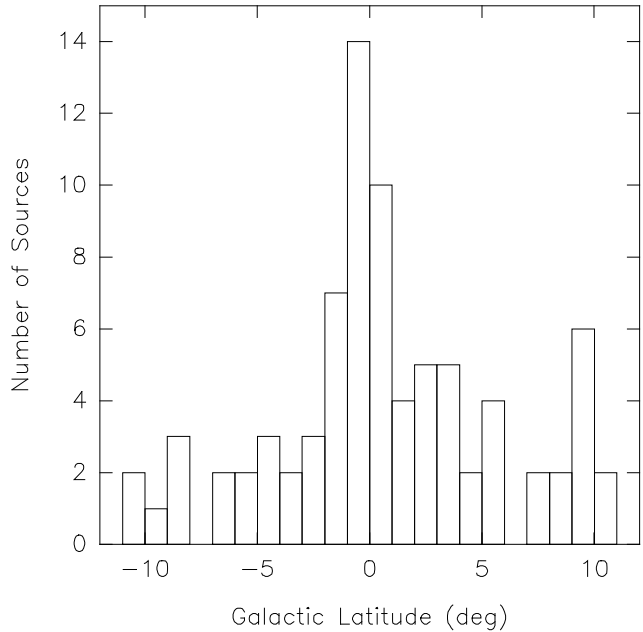, width=0.7\hsize}}
\caption{Distribution with galactic latitude of those unidentified
gamma-ray sources in the 3EG catalog located at $|b|<10^{\circ}$.}
\end{figure}

One of the first things that we could ask about the group of
low-latitude gamma-ray sources is whether they are correlated with
the spiral arms of the Galaxy. The arms are where most stars are
formed and the regions where the galactic gas storage is
concentrated. In order to test the correlation, we need a tracer
for the spiral structure of the Galaxy. The usual tracers, in this
sense, are giant and bright HII regions. We can use, then,
Georgelin \& Georgelin's (1976) catalog of the 100 brightest HII
regions to perform a correlation analysis with the sample of
low-latitude sources in the third EGRET catalog. When this is
done, we find that 32 out of 75 sources are positionally
overlapping HII regions. In order to quantify the statistical
significance of this number we can make numerical simulations of
random populations of galactic gamma-ray sources (subject to
adequate boundary conditions) using the code developed by Romero
et al. (1999) to produce synthetic gamma-ray source populations.
Through thousands of simulations, we find that the expected number
of chance coincidences is $13.2\pm2.9$. This implies a Poisson
probability of chance correlation of $5\times10^{-6}$, i.e. the
correlation is reflecting a physical relation with a confidence of
$\sim7\sigma$. The conclusion, consequently, seems to be that {\sl
there is a significant number of extreme Population I objects in
the parent population of low-latitude gamma-ray sources}. This
result was already suggested by previous correlation studies using
former gamma-ray catalogs presented by Montmerle (1979) and
Yadigaroglu \& Romani (1997), but the confidence only reaches
overwhelming levels when data from the third EGRET catalog are
used.

\section{Correlations with galactic objects}

The main mechanisms for gamma-ray production in a galactic
scenario are inverse Compton (IC) scattering of lower frequency
photons, relativistic bremsstrahlung and $\pi^0$-decays from
hadronic interactions. The common feature of all these mechanisms
is that they require the presence of a population of relativistic
particles (electrons or positrons in the first two cases, protons
or ions in the latter). Consequently, if we look for gamma-ray
production sites in the Galaxy, we should look at sites where
charged particles can be efficiently accelerated up to high
energies.

Basically, we have two types of scenarios where particles can be
accelerated up to the required relativistic energies: 1) large
sites where the acceleration is mediated by strong shock fronts in
a first order, diffusive process, and 2) compact objects with very
strong electromagnetic fields where the acceleration occurs in a
single step. The first type of acceleration is expected to occur
in supernova remnants (SNRs) and also at the strong shock that
could be formed near very massive stars endowed with strong
supersonic winds. The second type of acceleration should operate
in pulsars and accreting black holes, where strong magnetic fields
should be anchored in the surrounding accretion disks.

It is natural, then, to look for positional correlations between
gamma-ray sources and galactic objects like SNRs, early-type
stars, and OB associations (which are considered as pulsar
tracers), and this has been done in the past as we have briefly
mentioned in the Introduction. Regarding the sources in the third
EGRET catalog, Romero et al. (1999) have performed a correlation
analysis finding out that there is a suggestive number of spatial
coincidences with Wolf-Rayet and Of stars, SNRs and OB
associations. The probabilities of pure chance superposition are
moderate ($<10^{-2}$) for stars and quite negligible for remnants
and star forming regions ($<10^{-5}$). In the next sections we
shall discuss these potential gamma-ray emitters in more detail,
with emphasis in the case of stars.

\section{Stars}

Wolf-Rayet (WR) stars are very massive objects that have burnt
their hydrogen and have entered in the final phases of their
evolution. These stars present very strong supersonic winds with
velocities of several thousands of km/s. The mass loss rate is as
high as $10^{-4}$ $M_{\odot}$/yr. Of stars are also early-type
stars with strong winds; they are thought to be the progenitors of
WR stars and, although their winds are not so strong, they have
mass loss rates that can be $10^{9}$ times higher than those
presented by stars like the Sun.

The energy losses experienced by all these stars through their
winds have a great impact in the surrounding interstellar medium.
The winds swept up the ambient gas creating low-density cavities
around the stars (e.g. Benaglia \& Cappa 1999). A strong shock
front is formed at the contact layer between the wind and the
outer, colder ISM. These shocks are expected to accelerate
particles up to high-energies through Fermi mechanism (e.g. V\"olk
\& Forman 1982). If a source of UV photons or a cloud are present
near the acceleration site, significant amounts of gamma-rays
could be produced through IC scattering, bremsstrahlung or
hadronic interactions (Benaglia et al. 2000).

\begin{figure}[ht]
\centering
\resizebox{9cm}{!}{\includegraphics{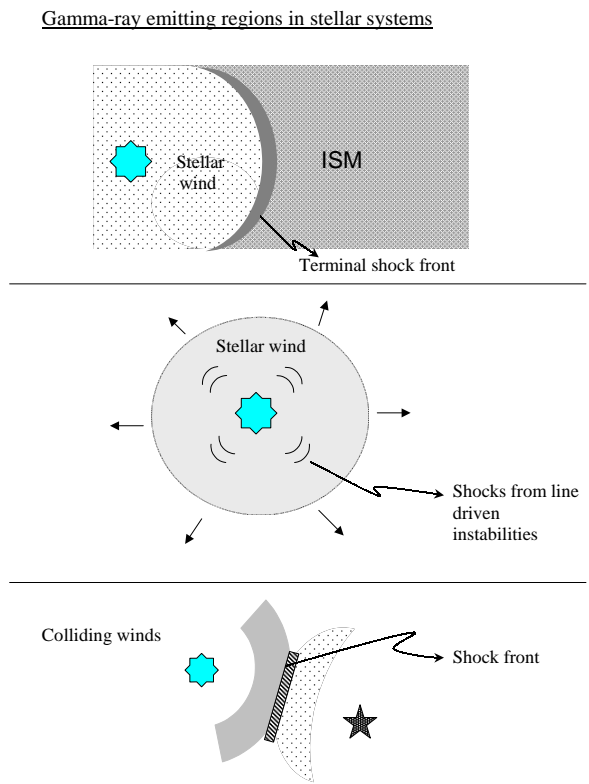}}\hfill
\caption{Sketch illustrating the different regions where
gamma-rays can be produced in a stellar system. Top panel:
Terminal shock region. Middle panel: The unstable base of the
wind. Lower panel: Colliding winds region in a binary system.}
\end{figure}

The winds of early-type stars are radiatively driven by absorption
in spectral lines and are prone to undergo instabilities that can
grow up to form strong shocks at the base of the outflow (Lucy \&
White 1980, Lucy 1982). These shocks can efficiently accelerate
both electrons and protons up to energies of about a few GeV
(White 1985), and these particles, through interactions with
stellar photons and ions, could yield gamma-ray emission in
EGRET's energy range (Benaglia et al. 2000).

Another site where stellar high-energy emission can be generated
is the colliding winds region in binary systems formed by two
massive stars (Eichler \& Usov 1993). Electrons are accelerated at
the strong shock formed by the winds collision, as evidenced by
the clear detection of non-thermal radio emission from the region
between stars in several systems (e.g. Contreras et al. 1997).
Gamma-rays should be produced by Comptonization of stellar photons
in this region.

\begin{table*}[ht]
\caption[]{WR stars spatially correlated with 3EG sources.}
\begin{flushleft}
\begin{tabular}{l l c c c c c l}
\hline \noalign{\smallskip} $\gamma$-Source & Star &
$\Delta\theta$ & $r$ & $v_{\infty}$ & $\log$\.M & NT \cr (3EG J) &
& (deg) & (kpc) & (km s$^{-1}$) &(M$_{\odot}$ yr$^{-1}$) &
emission
\\ \noalign{\smallskip} \hline \noalign{\smallskip} $0747-3412$ &
 WR 9 (B) & 0.37 & 2.35 & 2200 & $-4.2$ & \\
$1102-6103$ & WR 34 & 0.48 & 9.50 &1200 & $-4.5$&\\
    & WR 35     &  0.30 & 9.58 & 1100 & $-4.3$ &\\
    & WR 37     &  0.45 & 2.49 & 2150 & $<-4.1$&\\
    & WR 38    &  0.45 & 1.97 & 2400& $<-4.2$ &\\
    & WR 39     &  0.51 & 1.61 & 3600 & $<-4.0$
& Yes \\ $1655-4554$ & WR 80 & 0.59 & 4.40 & 2000 & $-4.1$&\\
$2016+3657$ & WR 137 (B)& 0.44 & 1.82 & 1900 & $-4.5$& \\ & WR 138
& 0.50 & 1.82 & 1500& $-4.7$&\\ $2021+3716$ &  WR 142& 0.15 & 0.95
& 5200 & $<-4.7$& \\ $2022+4317$ & WR 140 (B)& 0.64 & 1.34 & 2900&
$-4.1$&Yes
\\ \noalign{\smallskip} \hline
\end{tabular}
\end{flushleft}
\end{table*}

\begin{table*}[ht]
\caption[]{Of stars spatially correlated with 3EG sources.}
\begin{flushleft}
\begin{tabular}{l l c c c c c l}
\hline \noalign{\smallskip} $\gamma$-Source & Star &
$\Delta\theta$ & $r$ & $v_{\infty}$ & $\log$\.M & NT \cr (3EG J) &
& (deg) & (kpc) & (km s$^{-1}$) &(M$_{\odot}$ yr$^{-1}$) &
emission
\\
\noalign{\smallskip} \hline \noalign{\smallskip} $0229+6151$ &  HD
15629  &  0.30 &1.9 &2900 & $-5.8$ &\cr $0634+0521$ & HD 46150 &
0.64 & 1.3 &2900 & $<-5.9$ &\cr  & HD 46223 &  0.63 & 1.6 & 2800 &
-5.8 & \cr $1410-6147$ & HD 124314 & 0.25 & 1.0 & 2400 & $-4.7$ &
Yes \cr $2033+4118$  & Cyg OB2 5 (B)& 0.27 & 1.8 & 1500 & $-4.43$
& Yes \cr & Cyg OB2 11&  0.24 &1.8  & 2500 & $-5.2$&\cr
\noalign{\smallskip} \hline
\end{tabular}
\end{flushleft}
\end{table*}

In Tables 1 and 2 we list those unidentified gamma-ray sources
that are positionally coincident with WR and Of stars,
respectively. A ``B" letter marks those stellar systems that are
confirmed binaries. From left to right we provide the name of the
gamma-ray source in the 3EG catalog, the star name, angular
separation between the star and the best estimated position of the
3EG source, distance to the star, terminal wind velocity and mass
loss rate. We also indicate (in the last column) whether
non-thermal (NT) radio emission has been detected. The presence of
synchrotron radiation is important because it reveals the
existence of a population of relativistic electrons that could be
also responsible for the gamma-ray emission. In Figure 3 we show
the non-thermal radio contours of the southern Of star HD 124314
observed with the Australia Telescope Compact Array (ATCA) by
Benaglia et al. (2001). Similar observations for the remaining
sources listed in Tables 1 and 2 are in progress. The source 3EG
2016+3657, positionally coincident with WR 138, has been recently
studied by Mukherjee et al. (2000) who suggest that the
counterpart is the blazar B2013+370.

\begin{figure}[ht]
\centerline{\epsfig{file=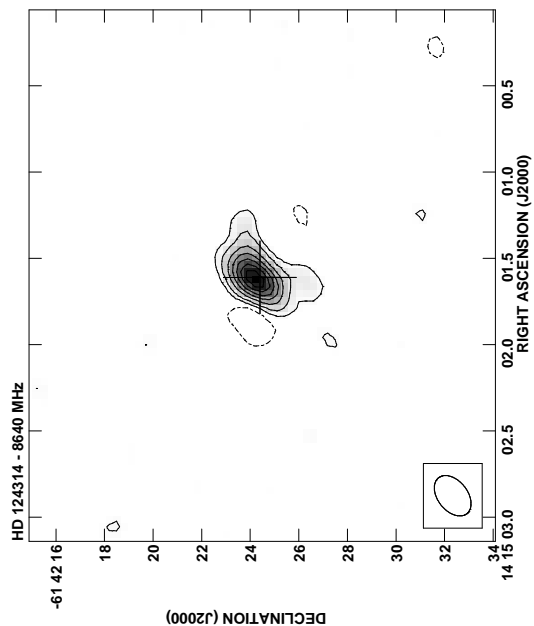, width=0.7\hsize}}
\caption{Nonthermal radio emission from HD124314 (from Benaglia et
al. 2001). The same population of relativistic electrons that
produces the observed synchrotron flux should produce gamma-rays
through Comptonization of UV stellar photons.}
\end{figure}

\begin{table}[ht]
\caption[]{Expected gamma-ray production in Cygnus OB2 No. 5 in
the EGRET energy range (from Benaglia et al. 2000)}
\begin{center}
\begin{tabular}{ l | l | c | c }
 \hline &&&\\
 Region    &  Mechanism      & Expected luminosity & Observed  luminosity \\
           &                 &  (erg/s)   &    (erg/s)   \\
&&&\\ \hline &&&\\
           & IC scattering & $\sim8\times 10^{34}$& \\
&&&\\ \cline{2-3} &&&\\ Winds collision     & Bremsstrahlung &
$\sim3.4\times 10^{30}$& \\ &&&\\ \cline{2-3} &&&\\
           & $\pi^0$-decay &   $\sim5.2\times 10^{24}$& \\
&&& $\sim2.4\times 10^{35}$ \\ \cline{1-3} &&&\\ Terminal shock &
 $\pi^0$-decay   &    $\sim2.3\times 10^{32}$& \\ &&&\\ \cline{1-3} &&&\\
           & IC scattering & -- &\\
Base of the wind &&&\\ \cline{2-3} &&&\\
          &  $\pi^0$-decay$^1$  &    $\sim5\times 10^{34}$& \\
&&&\\ \hline \multicolumn{4}{l} {1: See also White \& Chen
(1992)}\cr
\end{tabular}
\end{center}
\end{table}

The information on the relativistic electronic population obtained
with interferometric radio observations can be used, along with
the available information on the stellar parameters, to estimate
the expected gamma-ray luminosity of individual, well-studied,
stellar systems. Such work has been recently made by Benaglia et
al. (2000) for the particularly interesting case of Cyg. OB 2 No.
5. This system is formed by three stars: an O7 Ia + Ofpe/WN9
contact binary and a B0 V star in a larger orbit. Benaglia et al.
estimated the gamma-ray production in the colliding winds region,
the terminal shock of the dominant wind, and the unstable zone at
the base of the wind. Their results indicate that the high-energy
emission is dominated by IC radiation from the colliding winds and
$\pi^0$-decays from the inner wind region near the primary star.
The expected luminosities are shown in Table 3 and can be
considered as representative for similar systems.

Both INTEGRAL and GLAST missions will provide new and valuable
elements to test our ideas about how gamma-rays can be generated
in stars. In particular, the Imager on Board INTEGRAL Satellite
(IBIS), a coded masked detector designed to produce sky images in
the 20 KeV -- 10 MeV band with an angular resolution of 12
arcminutes, is expected to detect the IC flux density in a few,
nearby stars that have been already found to be non-thermal radio
sources. This instrument will also have a good spectral capability
that could be used to differentiate stars where the high-energy
emission comes from particles accelerated at a single shock (e.g.
the shock formed by winds collision) from those where the emission
is produced by particles accelerated at multiple shocks at the
base of the wind (e.g. Chen \& White 1991). The main limitations
of IBIS are the sensitivity (long integrations of $\sim 10^6$
seconds will be necessary to detect even the nearest stars) and
the fact that $\pi^0$-decays produce gamma-rays of energy well
above instrument's energy range. GLAST, on the contrary, will
detect this latter emission. Its improved sensitivity will allow
to observe thousands of sources, including hundreds of stars. The
determination of the spectral features and variability
characteristics of these stars, along with new and more accurate
measurements of their radio properties, will lead to a radical
advancement in our understanding of high-energy stellar processes.

\section{Supernova remnants}

Torres et al. (these proceedings) discuss in detail the case for
the association of SNR and 3EG sources. If we exclude sources that
are suspected to be artifacts, we have that 19 out 75 low-latitude
EGRET detections are positionally coincident with SNR in Green's
(2000) catalog (see also Romero et al. 1999). The probability of
all these associations being the mere effect of chance is quite
negligible: $\sim10^{-5}$. However, the possibility that many
sources result from the compact object left by the supernova
explosion and not from the remnant itself cannot be ruled out.

In several cases, however, the gamma-ray emitter seems to be the
SNR. This occurs when the remnant is interacting with nearby and
dense molecular clouds. In these cases, as discussed by Aharonian
et al. (1994) and Aharonian \& Atoyan (1996), the probability of
detection is dramatically increased. Some particular cases have
been studied in detail by Esposito et al. (1996) and Combi et al.
(1998, 2001). Cosmic rays are accelerated at the remnant shock
front and transported into the cloud by convection. There,
gamma-rays are produced by $\pi^0$-decays from $pp$ interactions
and relativistic bremsstrahlung. Additional contributions from IC
scattering can be also important. The spectral shape at low and
high energies can be used to infer the energy distributions of
both hadrons and leptons in the remnant. Multifrequency studies of
these SNRs can shed light on the details of the acceleration
process of galactic cosmic rays at large, strong shock fronts (see
Gaisser et al. 1998 for further details).

\section{OB associations}

OB associations are considered as tracers of young pulsar
concentrations (Kaaret \& Cottam 1996). Taking into account the
distribution of pulsar transverse speeds derived by Lyne \&
Lorimer (1994), it has been estimated that a fraction of
$0.25-0.4$ of all young pulsars (ages less than $10^{5}$ yr) can
be found within $1^{\circ}$ from the center of OB associations
(Kaaret \& Cottam 1996).

A correlation analysis using the 3EG catalog and Mel'nik \&
Efremov's (1995) OB associations catalog indicates that 26
low-latitude gamma-ray sources are superposed to OB associations.
This result is about $5\sigma$ away from what is expected by
chance (probability of all coincidences being the effect of chance
$\sim10^{-5}$).

\begin{figure}[ht]
\centerline{\epsfig{file=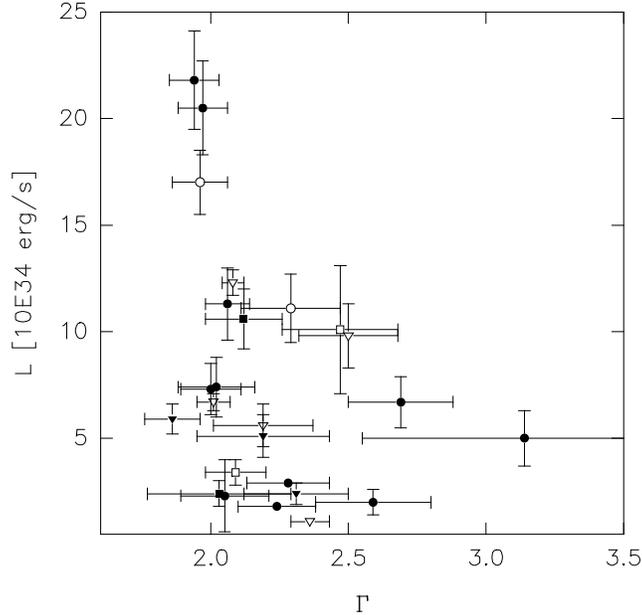, width=0.7\hsize}}
\caption{Expected luminosity versus high-energy spectral index for
unidentified 3EG sources that are correlated with OB
associations.}
\end{figure}

The idea that most of the gamma-ray sources in this group are
pulsars is supported by the particularly hard spectra presented by
them. All known gamma-ray pulsars are young objects with spectral
indices smaller than 2.15 (Crab's) and a trend for spectral
hardening with characteristic age (Fierro et al. 1993). In Figure
4 we show a plot of the expected luminosity of the gamma-ray
sources (assuming for them the distance to the corresponding OB
associations) versus the spectral index $\Gamma$ of the
high-energy emission ($F(E)\propto E^{-\Gamma}$). In this plot
pulsars should be located towards the left side of the panel,
precisely where most sources are concentrated. Three sources
differentiate from the rest, with high luminosities, similar to
the luminosity of Vela pulsar.

Recent modeling of gamma-ray pulsar populations by Zhang et al.
(2000) also corroborates that most sources positionally associated
with OB star forming regions are probably pulsars. The simulations
made by these authors, based on the outer gap model proposed by
Zhang \& Cheng (1997), shows that a mixed population of
radio-quiet and young radio pulsars can account for most (although
not all) associations found by Romero et al. (1999) using the the
3EG catalog.

\section{Uncorrelated sources}

The correlation analysis made by Romero et al. (1999) showed that
there is a significant number of low-latitude gamma-ray sources
for which there is no positional superposition with any known
galactic object thought to be capable of producing high-energy
gamma rays. To clarify the nature of this group of sources is,
perhaps, one of the most interesting problems aroused by EGRET.

A comparative analysis of the gamma-ray variability properties of
these sources could provide some clues on their origin. Gamma-ray
pulsars are expected not to vary over relatively short timescales,
so the identification of clearly variable galactic sources could
lead to the discovery of a new population of gamma-ray emitters in
the Galaxy (Tavani et al. 1997).

A variability analysis for the group of sources without positional
correlation has been performed by Romero et al. (2000) and Torres
et al. (2000). They have used a statistical method widely applied
to blazar studies (Romero et al. 1994). Although this method is
applied to the catalog data, it is free of several problems that
affect similar methods (see Reimer, this proceedings, and Tompkins
1999). Basically, a variability index $I=\mu_{\rm s}/<\mu>_{\rm
p}$ is introduced, where $\mu_{\rm s}=\sigma/<F>$ is the
fluctuation index of the gamma-ray source and $<\mu>_{\rm p}$ is
the averaged fluctuation index of all known gamma-ray pulsars
(which are usually considered as a non-variable population).
Sources with $I>1$ at $3\sigma$ or more are considered as possibly
variable ones ($1\sigma=0.5$). In the estimate of the fluctuation
index of known gamma-ray pulsars only data from the 3EG catalog
are considered. In the limits of very variable and non-variable
sources this method yields similar results to the most
comprehensive method applied by Tompkins (1999), who worked with
the raw data and took into account background fluctuations, source
contamination, and other sources of systematic error.

\begin{figure}[ht]
\centerline{\epsfig{file=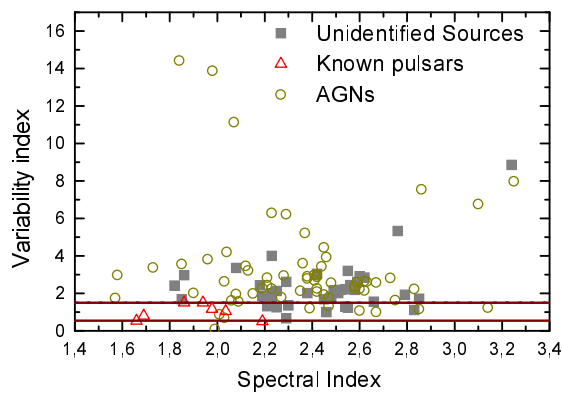, width=0.7\hsize}}
\vspace*{-5.5cm} \caption{Plot of variability index $I$ versus
high-energy spectral index $\Gamma$ for low-latitude 3EG sources
without positional correlation with potential galactic gamma-ray
emitters. Known gamma-ray pulsars are contained between the
horizontal lines. Gamma-ray AGNs are shown for comparison.}
\end{figure}

The basic results of the variability analysis of the low-latitude,
uncorrelated gamma-ray sources in the 3EG catalog are shown in
Figure 5 as a plot of variability index $I$ versus the high-energy
spectral index $\Gamma$. AGNs and pulsars are also shown in this
plot for comparison. Known pulsars are located between the two
solid lines at the bottom of the panel. Since they also have hard
spectra, they are concentrated towards the left side of the frame.
AGNs, on the contrary, mostly display high variability levels and
in many occasions steep spectra. It can be clearly seen from the
plot that there is a group of unidentified sources with a
behaviour similar to that presented by AGNs. The surface density
of these sources is, however, much higher than what could be
expected through an extrapolation of the high-latitude AGN density
towards the galactic plane. Actually, there seems to be a trend
among these sources in the sense that those with the highest
variability levels also present the steepest spectral indices.
These sources seems to form a population of galactic, variable
sources, of which GRO J1838-04 (Tavani et al. 1997) and 3EG
J1828+0142 (Punsly et al. 2000) are extreme examples.

\section{Variable gamma-ray sources in the Galaxy}

The nature of the variable low-latitude sources remains a mystery.
The large error boxes of the EGRET detections make very difficult
the identification of potential lower frequency counterparts. A
single radio field of $1^{\circ}$ around the centre of the EGRET
95 \% probability location contour can contain as many as 50
point-like weak radio sources, most of which are also of unknown
nature (Torres et al. 2000). Although only the improved angular
resolution of GLAST will allow to isolate the more promising
candidates for counterparts and lead to a final identification, we
can speculate about some possibilities on the nature of these
sources. Among the main ones, we can mention:
\begin{itemize}

\item {\sl Early-type stars with
strong winds}. In our study of the positional correlation of
sources in the 3EG catalog with massive stars we have only
included extreme stars like WR and Of stars. However, other O and
B stars can also be sources of gamma-ray emission strong enough as
to be detected by EGRET, especially if they are forming binary
systems (Eichler \& Usov 1993, Benaglia et al. 2000) or are
located in a very rich environment. The luminosity of these stars
should not be very high, say around a few times $10^{34}$ erg
s$^{-1}$ (Benaglia et al. 2000), so only nearby sources would be
detected. Variability is naturally expected in these stars due to
geometric changes during the orbital evolution of the binaries and
also due to the effect of wind instabilities. Stars are
interesting candidates to explain mid-latitude sources probably
linked to the nearby Gould belt (Gehrels et al. 2000).

\item {\sl Pulsars}. Pulsars, under certain circumstances, can be
variable sources. For instance, when a pulsar forms a
non-accreting binary system with a massive star, it could be
subjected to a changing UV photon bath from the star (e.g. Tavani
\& Arons 1997). Isolated pulsars also could be variable through
some kind of quake-driven activity.

\item {\sl Faint microquasars}. This hypothesis is supported by
the recent discovery of a persistent and faint microquasar (LS
5039) by Paredes et al. (2000, also these proceedings)
positionally correlated with the source 3EG 1824-1514. Other, yet
undetected microquasars could switch between high and low states
due to periodic accretion instabilities. Recent evidence seems to
show that LS 5039 displays short-term radio variability. If IC
gamma-ray emission is associated with the radio emission, it
should also be variable.

\item{\sl Isolated black holes accreting from the interstellar
medium}. Bondi-Hoyle accretion of diffuse matter onto black holes
of several solar masses can yield moderate gamma-ray luminosities
(Dermer 1997). If the hole is moving through an inhomogeneous
medium, variable emission could be observed.

\item{\sl Non-pulsating (NP) black holes}. Maximally rotating
Kerr-Newman black holes (see Punsly, these proceedings, and
references therein) can support a magnetosphere and, if they are
located in a low-density medium, could, in principle, be gamma-ray
sources (Punsly 1998a, b). Variability naturally results from jet
instabilities. A model of this kind has been recently applied to
the source 3EG J1828+0142 (Punsly et al. 2000).

\end{itemize}

Although this list does not exhaust all possibilities, it is
enough to show that a variety of candidates can be postulated and
that new observations of improved quality are required.

\section{Final remarks}

EGRET has confirmed the finding made by COS-B of a galactic
population of gamma-ray sources and has shown that this population
is not formed by a unique class of objects. Pulsars, supernova
remnants, and early-type stars seem to be capable, under certain
circumstances, to emit detectable gamma-rays. The possibility of
other, more strange, gamma-ray emitting objects in the Galaxy
remains open. The search for these objects will be one of the most
interesting and challenging tasks of high-energy astrophysics in
GLAST era.

\begin{acknowledgments}
The author is very grateful to Dr. Alberto Carrami\~nana for
supporting his travel to Mexico as well as to all members of the
Local Organizing Committee for a wonderful workshop. He also
wishes to express his sincere thanks to the members of GARRA team,
Paula Benaglia, Jorge Combi and Diego Torres, for many discussions
and fruitful collaborations on the topics covered in this review.
Comments by Isabelle Grenier are also gratefully acknowledged.
High-energy astrophysics with G.E. Romero is supported by ANPCT
through grant PICT 98 No. 03-04881 and Fundaci\'on Antorchas.
Additional support is provided by CONICET.
\end{acknowledgments}

\end{document}